\documentclass[twocol]{ametsocV6.1}

\usepackage[uncertainty-mode=separate]{siunitx}
\usepackage{booktabs}
\usepackage{bm}
\usepackage{amsmath} 
\usepackage[capitalise]{cleveref} 
\usepackage{amssymb} 
\usepackage{tikz}
\usetikzlibrary{shapes,arrows, arrows.meta, positioning, calc}

\title{Increasing NWP Thunderstorm Predictability Using Ensemble Data and Machine Learning}

\authors{Kianusch Vahid Yousefnia,\aff{a}\correspondingauthor{Kianusch Vahid Yousefnia, kianusch.vahidyousefnia@dlr.de} 
Tobias Bölle,\aff{a} 
Christoph Metzl,\aff{a} 
}

\affiliation{\aff{a}{Deutsches Zentrum für Luft- und Raumfahrt, Institut für Physik der Atmosphäre, Oberpfaffenhofen, Germany}
}

\abstract{
While numerical weather prediction (NWP) models are essential for forecasting thunderstorms hours in advance, NWP uncertainty, which increases with lead time, limits the predictability of thunderstorm occurrence. This study investigates how ensemble NWP data and machine learning (ML) can enhance the skill of thunderstorm forecasts.
Using our recently introduced neural network model, SALAMA 1D, which identifies thunderstorm occurrence in operational forecasts of the convection-permitting ICON-D2-EPS model for Central Europe, we demonstrate that ensemble-averaging significantly improves forecast skill. Notably, an 11-hour ensemble forecast matches the skill level of a 5-hour deterministic forecast. To explain this improvement, we derive an analytic expression linking skill differences to correlations between ensemble members, which aligns with observed performance gains. This expression generalizes to any binary classification model that processes ensemble members individually.
Additionally, we show that ML models like SALAMA 1D can identify patterns of thunderstorm occurrence which remain predictable for longer lead times compared to raw NWP output. Our findings quantitatively explain the benefits of ensemble-averaging and encourage the development of ML methods for thunderstorm forecasting and beyond.
}

\begin{document}

\maketitle

%
%
%
\statement
This study aims to improve thunderstorm forecasts, which are important for reducing the risks posed by hazards, such as lightning, heavy rain, and strong winds. Combining machine learning with an ensemble weather prediction model,  we found that averaging ensemble predictions greatly enhances forecast skill, making an 11-hour forecast as skillful as a 5-hour prediction without averaging. We also developed a formula explaining this improvement and demonstrated how ML can uncover patterns which make thunderstorms predictable for a longer period of time.  These results help improving existing tools for anticipating severe weather multiple hours ahead.
%
%

%

\section{Introduction}\label{sec:introduction}
Due to their accompanying hazards, such as lightning, heavy rainfall, large hail, and strong winds, thunderstorms constitute convective weather phenomena for which accurate forecasts are immensely beneficial to society and the economy. 
As the skill of nowcasting methods based on the extrapolation of remote sensing data quickly deteriorates in the course of \SI{1}{\hour} \citep{Leinonen2023}, thunderstorm forecasts multiple hours ahead rely on numerical weather prediction (NWP) \citep[e.g.][]{Simon2018, Brunet2019}. However, the prediction of convective phenomena which are very localized in space and time, such as thunderstorms, remains a significant challenge even for convection-permitting NWP models, particularly in terms of resolving their precise location and timing \citep{Roberts2008a, Weisman2008}. This issue arises from model grid spacings too coarse to resolve single-cell convection and insufficient sub-grid-scale parametrization of model physics, and is exacerbated by rapid error growth in time due the fundamentally chaotic nature of the Navier-Stokes equations \citep{Lorenz1969, Palmer2017, Craig2021}. As a result of NWP uncertainty, the skill of thunderstorm forecasts based on NWP drops for increasing lead times, which ultimately limits the practical predictability of thunderstorm occurrence.

Valuable information on the NWP uncertainty of a forecast is provided by ensemble models, and can be harvested to improve on forecasts by deterministic models \citep{Richardson2000, Zhu2002,Schwartz2017}.  Indeed, previous studies \citep{Schwartz2015, Loken2017} show that combining member-wise severe weather forecasts via ensemble-averaging improves forecast skill,  especially on mesoscale length scales relevant for deep convection \citep{Sobash2016}. On the other hand, we are not aware of studies that could quantitatively explain the increase in skill between ensemble-averaged and deterministic thunderstorm forecasts.

To identify thunderstorm occurrence in their NWP data, \cite{Sobash2011} and \cite{Loken2017} used the 2--\SI{5}{\kilo\meter} updraft helicity.  Traditional surrogates for convective environments and thunderstorms are motivated by a combination of physical considerations and empirical expertise. In addition to updraft helicity, examples include convective available potential energy \citep[e.g.,][]{Kaltenboeck2009, Taszarek2021}, or simulated radar reflectivity \citep[e.g.,][]{Kain2008, Kerr2025}. Recently, machine learning (ML) methods based on artificial neural networks have turned out to be powerful at systematically processing multi-variable NWP output \citep[e.g.,][]{Ukkonen2017,Kamangir2020,Geng2021,Jardines2024}. These methods are adjusted to pick up intricate non-linear patterns in the NWP output by being presented with a data set of past NWP forecasts and corresponding observations of the weather phenomenon of interest, such as thunderstorm occurrence. Several studies suggest that ML-based thunderstorm forecasts are more skillful than forecasts based on traditional surrogates for thunderstorm occurrence \citep{Ukkonen2019,Sobash2020,Yousefnia2024}.

To our knowledge, ML models for identifying thunderstorm occurrence have all been constructed as single-member models, by which we mean that either they process each ensemble member of the underlying NWP forecast individually \citep{Jardines2024,Yousefnia2024}, or they were trained on deterministic NWP forecasts altogether \citep[e.g.,][]{Kamangir2020,Geng2021}. Based on the previously cited literature on the success of ensemble-averaging, we expect the skill of single-member models to increase when applying them to all members of the NWP model and evaluating the ensemble mean. The focus of this present work is to quantitatively establish the added benefits of \emph{(i)} ensemble-averaging, and of \emph{(ii)} applying an ML model for identifying thunderstorm occurrence instead of using a traditional surrogate.

This work builds on our recently introduced neural network model SALAMA 1D \citep[Signature-based Approach of Identifying Lightning Activity Using Machine Learning 1D,][]{Yousefnia2024b}. It is a single-member ML model which infers the probability of thunderstorm occurrence in forecasts of ICON-D2-EPS, a convection-permitting NWP ensemble model for Central Europe run operationally by the German Meteorological Service (DWD). We first show that by considering the ensemble mean over all members we can improve the skill of thunderstorm forecasts produced by SALAMA 1D for lead times up to (at least) \SI{11}{\hour}. Importantly, we show that the observed increase in skill compared to considering only a single NWP member (effectively a deterministic forecast) can be understood in a broader setting. We motivate why skill increases can be expected for general single-member models of any binary classification task and are guaranteed for a particular class of skill scores. For the Brier skill score, we work out an analytic expression for the increase in skill and validate it for the SALAMA 1D model using past NWP forecasts and observations of the ground-based lightning network LINET \citep{Betz2009}. Finally, we compare SALAMA 1D's lead-time dependence of skill with that of a simple surrogate model based on raw NWP output (without any ML) to study to what extent thunderstorm forecasts benefit from ensemble-averaging and ML individually. 

Our work may encourage severe-weather forecasters to apply the large range of recently developed single-member ML models directly to ensemble forecasts. The easily interpretable analytic expression for the corresponding increase in skill could provide guidance for ensemble NWP modelers on how to improve severe weather forecasting. Furthermore, we shall see that our results strengthen the role of multi-variable frameworks like ML in thunderstorm forecasting, while also exemplifying the benefit of post-processing raw NWP data with observational data.

We introduce our ML framework for ensemble thunderstorm forecasting in \cref{sec:methods}.  In \cref{sec:data}, we detail a pipeline for compiling data sets to eventually measure thunderstorm identification skill. We present our results in \cref{sec:results}, while summarizing and discussing their implications and possible research avenues in \cref{sec:conclusion}.

\section{ML ensemble thunderstorm forecasting}\label{sec:methods}
In this section, we introduce our ML model for identifying thunderstorm occurrence, as well as the ensemble NWP model on which it was trained, and detail the process of ensemble-averaging.

NWP ensemble systems estimate forecast uncertainty by producing multiple physically consistent forecasts. The variability between the ensemble members aims to reflect the NWP uncertainty in the initial conditions, model design, and boundary conditions. 
This work is based on forecasts by ICON-D2-EPS \citep{Zaengl2015,Reinert2020}, a convection-permitting NWP ensemble model operationally run by the German Meteorological Service (DWD). This NWP model has a horizontal resolution of $\sim\SI{2}{\kilo\meter}$, 65 vertical levels, and 20 ensemble members. 
In ICON-D2-EPS, the data assimilation system KENDA  \citep[Kilometer-scale Ensemble Data Assimilation,][]{Schraff2016} with a latent heat nudging scheme \citep{Stephan2008} combines current observations and a short-term forecast from the preceding data assimilation cycle to create a 40-member ensemble of consistent and statistically indistinguishable initial conditions, of which 20 members are propagated in time.
For each member, the values of certain NWP model parameters are sampled randomly within a given range to account for model uncertainty. Finally, model runs of a global ICON ensemble suite with a \SI{20}{\kilo\meter} nesting area over Europe provide hourly lateral boundary conditions for each member \citep{Reinert2020}.
While the spread in an NWP ensemble can by construction account only for statistical uncertainties and not for systematic model biases, we still expect the ensemble mean of thunderstorm forecasts to be more skillful than a deterministic forecast, as uncertainties and noise may in part average out. The main objective of this paper is to understand quantitatively the potential increase in skill. 

We exemplify our methodology of comparing ensemble-averaged thunderstorm forecasting skill to deterministic skill using our ML-based model SALAMA 1D \citep{Yousefnia2024b}, of which we now introduce its main aspects. We stress, however, that the emphasis of this work lies on our methodology, which generalizes to any other model that processes the NWP output of a single member and returns probabilities of thunderstorm occurrence (or any other weather phenomenon, for that matter).

SALAMA 1D is a neural network model set up to solve what is called a binary classification task in ML terminology: Considering the vertical profiles of $N_\text{f}=10$ atmospheric variables (\cref{tab:fields}) of a given ICON-D2-EPS member at $N_z$ vertical levels, it infers the corresponding probability of thunderstorm occurrence,
\begin{equation}
\text{SALAMA 1D:} \quad \mathbb{R}^{N_\text{f}\times N_z} \to (0,1) .
\label{eq:SALAMA1D}
\end{equation}
Neural networks model the functional relationship between input and output through multiple sequential executions of matrix multiplications, vector additions and nonlinear operations.  The matrices and vectors involved provide free parameters which are determined using a training set of input samples and the corresponding yes/no information whether a thunderstorm has occurred.
To the best of our knowledge, SALAMA 1D is the only thunderstorm identification model that evaluates vertical profiles of atmospheric variables rather than relying on multiple (single-level) thunderstorm surrogate variables, and demonstrate higher skill as a result. The ML model architecture is guided by two physical considerations. On the one hand, sparse connections encourage interactions at similar height levels while also significantly reducing parameter size. On the other hand, profiles are effectively shuffled in the vertical during training to prevent the model from learning non-physical patterns tied to the vertical grid structure of the NWP model. SALAMA 1D has been trained on NWP forecasts from two summers \citep[2021 and 2022; i.e., we use the SALAMA 1D-2022 variant from][]{Yousefnia2024b} with lead times between \SI{0}{\hour} and \SI{2}{\hour}. We emphasize that SALAMA 1D has been designed as a single-member model, by which we mean that, given the NWP forecast of a given ensemble member, SALAMA 1D provides the corresponding probability of thunderstorm occurrence. The model has been trained on randomly chosen ensemble members of ICON-D2-EPS.



Next, we describe the necessary details for combining SALAMA 1D forecasts via the ensemble mean.
ICON-D2-EPS forecasts include for each grid point $N_\text{e}=20$ sets of vertical profiles $\left(\bm{\xi}^{(k)}\right)_{k=1,\dots, N_\text{e}}$, with the $\bm{\xi}^{(k)}\in\mathbb{R}^{N_\text{f}\times N_z}$ originating from the individual ensemble members. We produce a single-valued thunderstorm forecast for a given grid point by applying SALAMA 1D to each set of vertical profiles separately, which yields member-wise probabilities $\left(p^{(k)}\right)_{k=1,\dots, N_\text{e}}$. We then compute the ensemble mean:
\begin{equation}
  \langle p \rangle = \frac{1}{N_\text{e}}\sum_{k=1}^{N_\text{e}}p^{(k)} = \frac{1}{N_\text{e}}\sum_{k=1}^{N_\text{e}}\text{SALAMA 1D}\left(\bm{\xi}^{(k)}\right)
  \label{eq:ensemble_mean}
\end{equation}
We refer to the application of SALAMA 1D to ensemble data in the above manner as
\begin{equation}
\text{SALAMA 1D-EPS:} \quad \mathbb{R}^{N_\text{e}\times N_\text{f}\times N_z} \to (0,1).
\label{eq:SALAMA1D-EPS}
\end{equation}


\section{Data preprocessing}\label{sec:data}
\begin{figure}[htbp]
\centering
\includegraphics[width=\columnwidth]{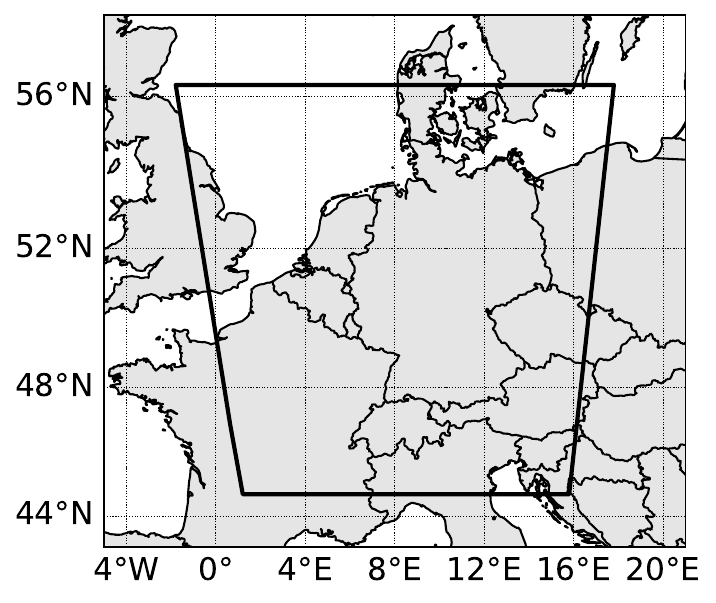}
\caption{Study region for this work. The polygon vertices, counterclockwise from the bottom-left, read: (\ang{44.7}N, \ang{1.2}E), (\ang{44.7}N, \ang{15.8}E), (\ang{56.3}N, \ang{17.8}E), (\ang{56.3}N, \ang{1.8}W)}
\label{fig:study_region}
\end{figure}

We aim to compare the lead-time dependence of classification skill for two evaluation modes of SALAMA 1D:
\begin{itemize}
    \item Evaluation of SALAMA 1D on a single ensemble member. This is how SALAMA 1D has been originally introduced \citep{Yousefnia2024b}. We refer to this evaluation mode as ``SALAMA 1D model''.
    \item Evaluation of SALAMA 1D on all ensemble members and computation of the ensemble mean (\cref{eq:ensemble_mean}). We refer to this evaluation mode as ``SALAMA 1D-EPS model''.
\end{itemize}
For this purpose, we compile data sets for testing the skill of our models for NWP forecasts with fixed lead times. We generate a separate data set for each lead time from \SI{0}{\hour} to \SI{11}{\hour}. Each data set consists of $N=\num{e5}$ pairs $(\bm{\xi},y)$, where $\bm{\xi}\in\mathbb{R}^{N_\text{e}\times N_\text{f}\times N_z}$ denotes the NWP data of all ensemble members for some random grid point in the study region (\cref{fig:study_region}) and some time during the testing period (\cref{sec:data}\ref{sec:NWP_data}), while the label $y\in\{0, 1\}$ denotes whether the given forecast $\bm{\xi}$ is associated with thunderstorm occurrence or not (1: ``thunderstorm occurrence'', 0: ``no thunderstorm occurrence''). To test the skill of a given ML model using a test set, we will later apply the model to all input samples $\bm{\xi}$ and compare the resulting model output probabilities with the labels $y$ via a skill score, such as the Brier skill score.

To assure a fair comparison, we evaluate both SALAMA 1D models on the same data sets: For SALAMA 1D-EPS, we can use the data set from above directly using \cref{eq:SALAMA1D-EPS}, since $\bm{\xi}$ contains the NWP forecasts of all ensemble members. For SALAMA 1D, we restructure the data set so that each ensemble member is treated as an independent sample, allowing us to apply \cref{eq:SALAMA1D} to a total of $N_\text{e} \times N = \num{2e6}$ pairs. The next sections provide details on the NWP data $\bm{\xi}$ and observations $y$.

\subsection{NWP data}\label{sec:NWP_data}
The operational runs of ICON-D2-EPS are initialized daily every \SI{3}{\hour}, starting at 0000~UTC, and produce hourly forecasts. We collect the forecasts of all runs from July and August, 2023, up to a lead time of \SI{11}{\hour}. We use the forecasts for which the target time falls on an even day of the month for test set compilation, since the odd days have been used for monitoring training progress of SALAMA 1D \citep{Yousefnia2024b}. In addition, adopting the approach from \cite{Yousefnia2024b}, we let the days start at 0800~UTC to reduce even further the risk of correlations between our test sets and the validation set used during the training of SALAMA 1D.

To compile a test set with NWP forecasts of a fixed lead time, we sample for each pair $(\bm{\xi},y)$ a random grid point from the study region and a random date from the testing period. 
The time of the day is randomly drawn from the full hours for which a forecast of the given lead time is available. For instance, the available hours of the day according to the DWD initialization schedule for \SI{1}{\hour} forecasts are 0001~UTC, 0004~UTC, 0007~UTC, and so on. 
Furthermore, we keep all ensemble members.

Therefore, a sample $\bm{\xi}$ has the shape $(N_\text{e}, N_\text{f}, N_z)=(20, 10, 65)$ and comprises the vertical profiles of the atmospheric variables listed in \cref{tab:fields}. 
\begin{table}[htbp]
\caption{ICON-D2-EPS variables used in this study.}
\label{tab:fields}
\begin{tabular}{rl}
\toprule
ICON variable & Description\\
\midrule
U & Zonal wind speed\\
V & Meridional wind speed \\
T & Temperature \\
P & Pressure \\
QV & Specific humidity \\
QC & Cloud water mixing ratio \\
QI & Cloud ice mixing ratio \\
QG & Graupel mixing ratio \\
CLC & Cloud cover \\
W & Vertical wind speed \\
\bottomrule
\end{tabular}
\end{table}

\subsection{Lightning observations}\label{sec:lightning}
In order to associate the NWP data for a given grid point and time with thunderstorm occurrence in the atmosphere, we utilize observation data from the lightning detection network LINET \citep{Betz2009}. Lightning observations of this source were also used in the training of SALAMA 1D due to their high and uniform detection efficiency ($\geq \SI{95}{\percent}$) and spatial accuracy ($\SI{150}{\meter}$).

Following \citet{Ukkonen2019, Yousefnia2024b}, we consider a thunderstorm to occur at $(x,t)$ if a flash of lightning is detected at any $(x_l,t_l)$ with
\begin{equation}
\left\| \bm{x} - \bm{x}_l \right\| < \Delta r \quad \text{and}\quad |t-t_l| < \Delta t.
\end{equation}
Here, $\left\| \cdot \right\|$ denotes the great-circle distance between $\bm{x}$ and $\bm{x}_l$ on a perfect sphere of the Earth with a radius of \SI{6371.229}{\kilo\meter}, as assumed in the ICON-D2-EPS model \citep{Reinert2020}.
We set the spatiotemporal thresholds to the same values used for training SALAMA 1D, i.e., $\Delta r = \SI{15}{\kilo\meter}$ and $\Delta t = \SI{30}{\minute}$. These thresholds result in a relative frequency $g=1.93^{\,+\,0.23}_{-0.24}\times 10^{-2}$ of thunderstorm occurrence (sample climatology), the uncertainty showing the symmetric \SI{90}{\percent} confidence interval \citep{Yousefnia2024b}.

\section{Results}\label{sec:results}
In \cref{sec:results}\ref{sec:EPS_skill}, we report on an increase in skill of SALAMA 1D-EPS with respect to SALAMA 1D. In particular, we derive an analytic expression for the difference in skill and show that the expression is consistent with measured difference in skill.
In \cref{sec:results}\ref{sec:NWP_skill}, we compare the skill decay of our ML model as a function of lead time to a simple benchmark model based on raw NWP output without any ML-based corrections.

\subsection{Benefit of ensemble data}\label{sec:EPS_skill}
\begin{figure*}[htbp]
\centering
\includegraphics[width=\textwidth]{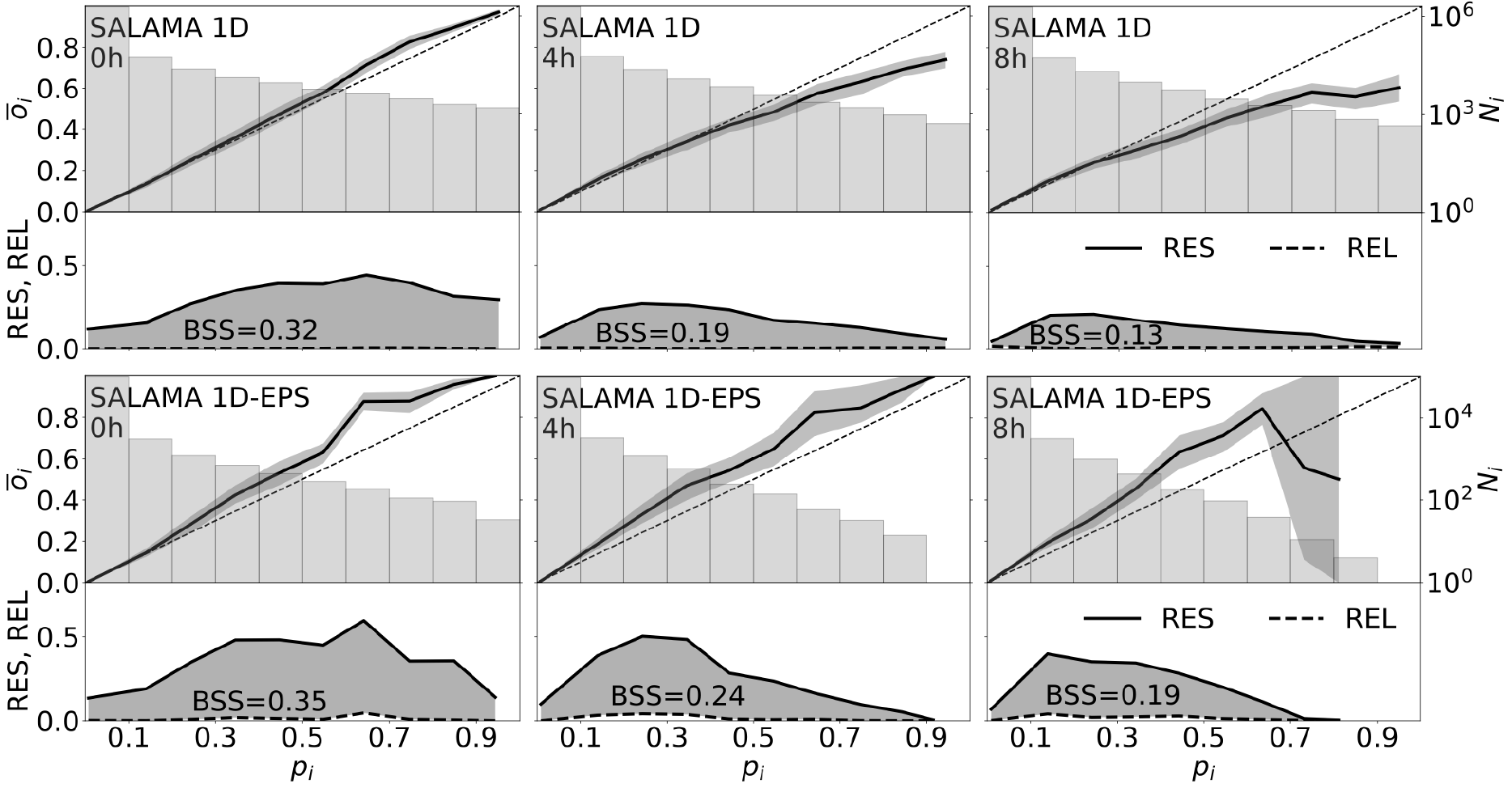}
\caption{Reliability diagrams and bin-wise reliability and resolution for SALAMA 1D (upper panels) and SALAMA 1D-EPS (lower panels) for the lead times \SI{0}{\hour} (left), \SI{4}{\hour} (middle), \SI{8}{\hour} (right). Shaded bands around the calibration functions denote uncertainties on a symmetric \SI{90}{\percent} confidence interval. Uncertainties are obtained from \num{e4} block bootstrap resamples, with day-wise block resampling.}
\label{fig:reliability_diagrams}
\end{figure*}

To compare SALAMA 1D and SALAMA 1D-EPS quantitatively and as a function of lead time, we use the test sets introduced in \cref{sec:data} to produce reliability diagrams \citep{Broecker2007,Wilks2011}.  A reliability diagram is constructed by first partitioning the range $(0,1)$ of model probabilities into $N_\text{b}$ equidistant bins and assigning each element in a given test set to a bin according to the element's model probability. For each bin $i=1, 2, \dots, N_\text{b}$, we retain the bin-averaged model probability $p_i$, the observed frequency $\overline{o}_i$ of thunderstorm occurrence, and the number $N_i$ of elements per bin. A reliability diagram displays the calibration function ($\overline{o}_i$ against $p_i$) and the refinement distribution ($N_i$ against $p_i$). The calibration function quantifies whether model probabilities exhibit reliability; i.e., whether they are consistent with observed relative frequencies.  The refinement distribution can be used to assess resolution, which describes the capability of a model to output well-calibrated probabilities larger than climatology. In \cref{fig:reliability_diagrams}, we show the reliability diagrams of SALAMA 1D and SALAMA 1D-EPS for the lead times \SI{0}{\hour}, \SI{4}{\hour}, and \SI{8}{\hour}, the upper half of each panel showing the corresponding calibration function and refinement distribution. SALAMA 1D is perfectly calibrated for the $\SI{0}{\hour}$ forecasts, outperforming SALAMA 1D-EPS, at least for high model probabilities. The higher reliability of SALAMA 1D is not surprising since this model was explicitly trained by loss-function optimization to output reliable single-member probabilities. The model was \emph{not} optimized for ensemble-averaged model probabilities to be reliable. Therefore, it is remarkable that SALAMA 1D-EPS produces reasonably reliable forecasts anyway. For longer lead times, both models show a similar degree of reliability.

As for resolution, it can be difficult to compare the two models solely from their refinement distributions, especially when the two models in question are similarly skillful. To assess resolution (and reliability) more easily, we consider bin-wise contributions to resolution and reliability, as introduced in \cite{Yousefnia2024}:
\begin{align}
\text{RES}_i &=  \frac{1/\Delta p}{g(1-g)}\frac{N_i}{N} (p_i - g)^2, \label{eq:res}\\
\text{REL}_i &=  \frac{1/\Delta p}{g(1-g)}\frac{N_i}{N} (p_i - \overline{o}_i)^2, \label{eq:rel}
\end{align}
where $\Delta p = 1/N_\text{b}$ denotes bin width. Note that resolution is a positively-oriented (``the higher, the better'') measure of skill, while reliability is negatively-oriented.  According to \cite{Murphy1973}, the area enclosed by $\text{RES}_i$ and $\text{REL}_i$ as a function of $p_i$ is equivalent to the Brier skill score (BSS) and, hence, a positively-oriented measure of skill. We will provide the formal definition of BSS later, in \cref{eq:BSS_BS}.

The lower half of each panel in \cref{fig:reliability_diagrams} displays the bin-wise reliability and resolution contributions to BSS. Inspection of the enclosed areas reveals that even though SALAMA 1D-EPS scores worse than SALAMA 1D in terms of reliability, its resolution is higher across all lead times, which yields a higher BSS. As lead time increases, both models' skill drops; SALAMA 1D-EPS, however, consistently outperforms SALAMA 1D in terms of BSS. Note that the higher skill results mainly from larger contributions to resolution for modest probabilities lower than \num{0.5}. Conversely, the contribution to skill from probabilities close to \num{1} are actually smaller for SALAMA 1D-EPS than they are for SALAMA 1D. This illustrates qualitatively how ensemble-averaging increases skill: The ensemble mean smooths out the rare high-probability predictions of individual members in favor of a less confident but overall more skillful averaged forecast. 

\begin{figure}[htbp]
\centering
\includegraphics[width=\columnwidth]{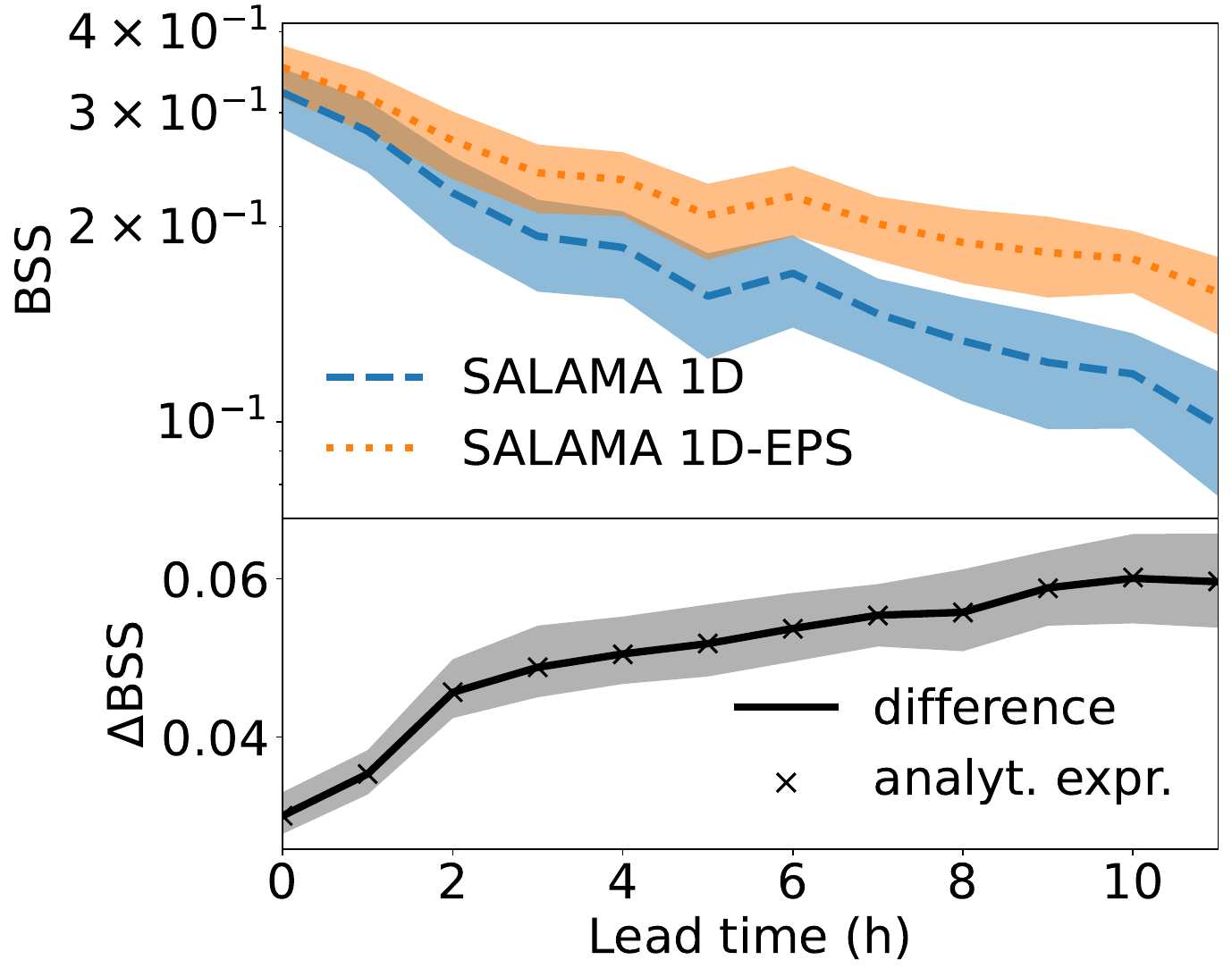}
\caption{Lead-time dependence of skill, quantified by the Brier skill score (BSS), of single-member forecasts (SALAMA 1D) and ensemble forecasts (SALAMA 1D-EPS) of thunderstorm occurrence. The lower panel shows the difference in skill, together with the prediction from the analytic expression \eqref{eq:diff_BSS}. Shaded bands correspond to sampling uncertainty for a symmetric \SI{90}{\percent} confidence interval. Uncertainties are obtained from \num{e4} block bootstrap resamples, with day-wise block-resampling.}
\label{fig:BSS_leadtime}
\end{figure}

For the remainder of this section, we study in more detail the lead-time dependence of BSS for the two models, which is shown in the upper panel of \cref{fig:BSS_leadtime}. SALAMA 1D-EPS outperforms SALAMA 1D significantly. Indeed, an \SI{11}{\hour}-forecast of SALAMA 1D-EPS is as skillful (in terms of BSS) as the \SI{5}{\hour}-forecast of SALAMA 1D.

Intuitively, ensemble-averaging leads to a more skillful forecast because we estimate the probability of thunderstorm occurrence based on a larger sample (of size $N_\text{e}=\num{20}$) than in the single-member case (sample size 1). We now derive this relationship more formally. It is instructive to introduce a probabilistic setting, as is common practice when investigating statistical properties of verification scores like BSS \citep{Broecker2007b,Bradley2008}. To this end, we introduce a tuple $(p, y)$ of random variables, where the discrete variable $y\in\{0,1\}$ denotes the thunderstorm occurrence ground truth and the continuous random variable $p\in(0,1)$ models the corresponding SALAMA 1D output. 
We denote the expected value and variance of $p$ by $\mathbb{E}[p] \equiv \overline{p}$ and $\text{Var}[p]\equiv \sigma^2$, respectively.  We first introduce the Brier score (BS), which is formally defined as the following expected value \citep{Brier1950, Wilks2011},
\begin{align}
\text{BS}_\text{single-mem} 
&= \mathbb{E} \left[(p-y)^2\right] \\
&= \sigma^2 + \overline{p}^2 - 2\mathbb{E}[py] + \mathbb{E}\left[y^2\right],
\label{eq:BS_det}
\end{align}
where we added the subscript ``single-mem'' to emphasize that this result holds for when evaluating a single ensemble member.
The Brier \emph{skill} score (BSS) from above is then related to BS by
\begin{equation}
\text{BSS}_\text{single-mem} = 1 - \frac{\text{BS}_\text{single-mem}}{\text{BS}_\text{ref}},
\label{eq:BSS_BS}
\end{equation}
where $\text{BS}_\text{ref} = \mathbb{E}\left[(g-y)^2\right]\equiv \kappa^2$ is a reference score using sample climatology $g\equiv\mathbb{E}[y]$.

\begin{figure}[htbp]
\centering
\includegraphics[height=0.49\columnwidth]{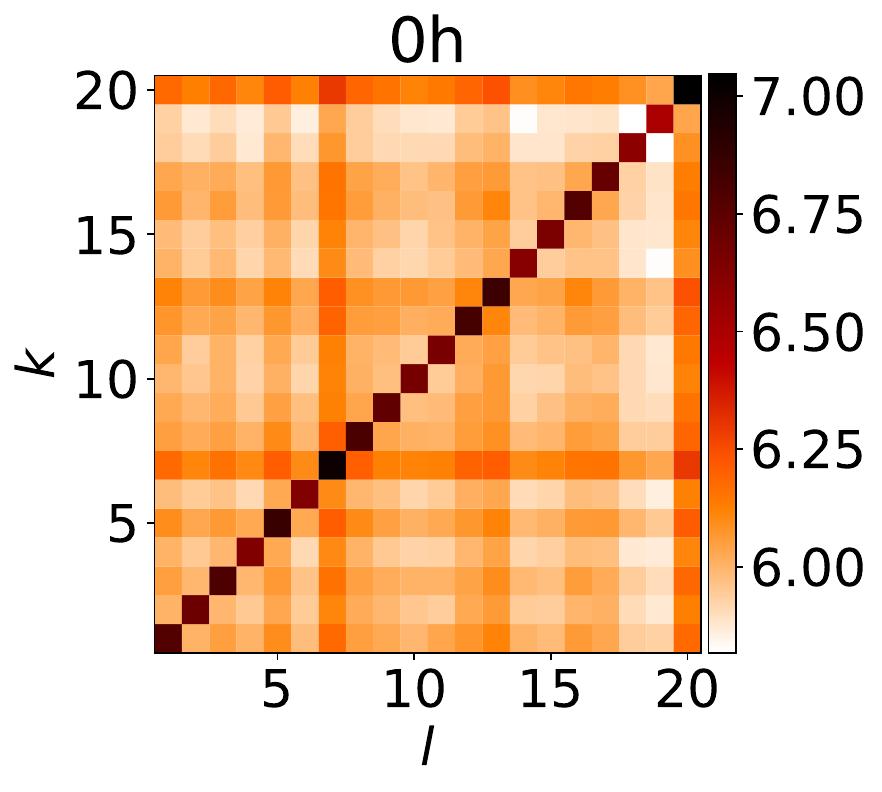}
\includegraphics[height=0.49\columnwidth]{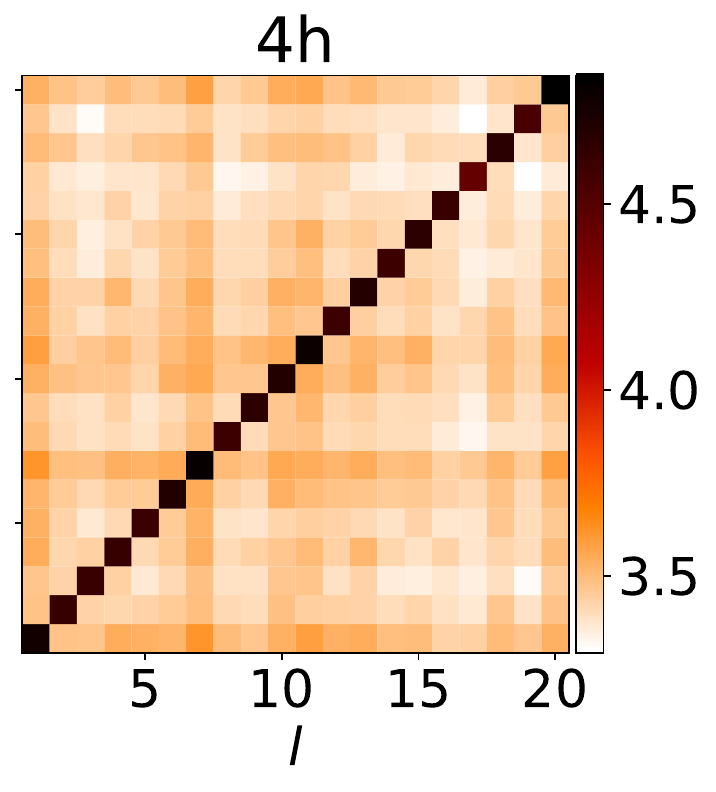}
\caption{Sample covariance matrix $\text{Cov}[p^{(k)}, p^{(l)}]/\num{e-3}$ of the member-wise probabilities $p^{(k)}, k = 1,\dots,N_\text{e}$, estimated for the test set of \SI{0}{\hour} lead time (left) and \SI{4}{\hour} lead time (right). If the members of the ensemble are exchangeable, the covariance matrix is fully determined by two numbers (one number for the diagonal entries of the matrix, one number for the off-diagonal entries), which is approximately the case.}
\label{fig:cov_structure}
\end{figure}

As for SALAMA 1D-EPS, we replace $p$ in the above framework by $N_\text{e}$ (possibly correlated) continuous random variables $p^{(k)}$, the arithmetic mean of which yields SALAMA 1D-EPS model output. We now make the crucial assumption that the random variables $p^{(k)}$ are exchangeable; i.e. their joint distribution is invariant under any permutation of the indices $\{1, \dots, N_\text{e}\}$. This assumption is motivated by the fact that we trained SALAMA 1D on all ensemble members without favoring any individual member. Moreover, the ensemble of perturbed initial conditions produced by the KENDA system in ICON-D2-EPS consists of statistically indistinguishable members (\cref{sec:methods}). Now, exchangeability implies that all $p^{(k)}$ have the distribution of $p$ from SALAMA 1D as their marginal distribution.  In addition, we have  $\mathbb{E}[p^{(k)}] \equiv \overline{p}$ and $\mathbb{E}[p^{(k)}y] = \mathbb{E}[py]$ for all $k$, and the covariance matrix of the $p^{(k)}$ takes on a particularly simple structure,
\begin{equation}
\text{Cov}\left[p^{(k)}, p^{(l)}\right] =
\begin{cases}
\sigma^2 & \text{if } k = l \\
\gamma & \text{otherwise,} \\
\end{cases} 
\label{eq:cov_structure}
\end{equation}
with $\sigma^2$ from above, and some number $\gamma\in\mathbb{R}$. In \cref{fig:cov_structure}, we exemplify the validity of parametrization \eqref{eq:cov_structure} for two test sets.

The BS for SALAMA 1D-EPS then reads:
\begin{align}
\text{BS}_\text{EPS} 
&= \mathbb{E}\left[\left(\frac{1}{N_\text{e}}\sum_{k=1}^{N_\text{e}}{p^{(k)}}-y\right)^2\right]\\
&= \frac{\sigma^2}{N_\text{e}} + \overline{p}^2+\frac{N_\text{e}-1}{N_\text{e}}\gamma - 2\mathbb{E}[py] + \mathbb{E}\left[y^2\right]
\label{eq:BS_EPS}
\end{align}
By subtracting \cref{eq:BS_det} from \cref{eq:BS_EPS}, we derive the BS difference $\Delta \text{BS} = \text{BS}_\text{EPS}-\text{BS}_\text{single-mem}$ between SALAMA 1D-EPS and SALAMA 1D, given by
\begin{equation}
\Delta \text{BS} = \frac{N_\text{e} - 1}{N_\text{e}}\left(\gamma - \sigma^2\right),
\label{eq:diff_BS}
\end{equation}
where we note that the case of uncorrelated members ($\gamma = 0$) is known in the ensemble ML community \citep{Abe2022}.
The reference score $\text{BS}_\text{ref}=\kappa^2$ depends only on the observations; hence, it is independent of the thunderstorm identification model under consideration. Therefore, it follows from \cref{eq:BSS_BS} that the difference $\Delta \text{BSS}$ between the two models is given by $-\Delta \text{BS}/\kappa^2$, which yields
\begin{equation}
\Delta \text{BSS}= \frac{N_\text{e}-1}{N_\text{e}\kappa^2}\left(\sigma^2-\gamma\right).
\label{eq:diff_BSS}
\end{equation}

Before discussing \cref{eq:diff_BSS} more thoroughly, we compare the lead-time dependence of $\Delta \text{BSS}$, as shown in the lower panel of \cref{fig:BSS_leadtime}, obtained through direct evaluation of the test sets with the values calculated using \cref{eq:diff_BSS}. To evaluate $\sigma^2$ and $\gamma$ for a given lead time, we used the corresponding test set to estimate the ensemble covariance matrix (\cref{fig:cov_structure}), and averaged over its diagonal entries (for $\sigma^2$), or off-diagonal entries (for $\gamma$).
We find excellent agreement between the two curves. Note, however, that since the test sets are used for the direct evaluation of $\Delta \text{BSS}$ and for computing $\sigma^2$ and $\gamma$, the two data series in the lower panel of \cref{fig:BSS_leadtime} are not independent evaluations of $\Delta \text{BSS}$. Nevertheless, their agreement justifies the assumptions that went into deriving \cref{eq:diff_BSS}, in particular, the $p^{(k)}$ being exchangeable.

Having validated \cref{eq:diff_BSS}, we discuss some immediate conclusions. First, notice that $\Delta \text{BSS}$ decreases when $\gamma$ approaches $\sigma^2$, which corresponds to ensemble members becoming increasingly correlated. \Cref{fig:cov_structure} shows that correlation between the members is quite high ( $\gamma \approx 0.85\sigma^2$ for lead times of \SI{0}{\hour}). This suggests that efforts to decrease inter-member correlations in the NWP ensemble \citep[e.g.,][]{Anderson2016,Necker2023,Morzfeld2023} are most promising for improving thunderstorm forecasting skill in terms of BSS.
Furthermore, as expected, $\Delta \text{BSS}$ increases for larger sample sizes $N_\text{e}$, the prefactor $(N_\text{e}-1)/N_\text{e}$ approaching \num{1}. According to \cref{eq:diff_BSS}, an ensemble size of $N_\text{e}=20$ already yields a factor of $19/20=0.95$, which would suggest that there is only little gain to be expected from increasing the size of the NWP ensemble. However, larger ensemble systems also reduce sampling errors in ensemble correlations, which could ultimately decrease $\gamma$ and increase $\Delta \text{BSS}$.

Finally, we acknowledge that our analysis focused on only one skill score, namely BSS. Note, however, that the qualitative trend seen in the upper panel of \cref{fig:BSS_leadtime} is equally recovered when using different skill scores. In particular, we checked this for the $F_1$-score, the critical success index, the equitable threat score, and the area under the precision-recall curve. The reason why we concentrated on BSS here is the mathematical tractability of this score, which allows for a closed-form expression of $\Delta \text{BSS}$.
Remarkably, the result that $\Delta \text{BSS} \geq 0$ can also be obtained from Jensen's inequality, which states that if a function $\varphi: \mathbb{R}\to\mathbb{R}$ is convex, then 
\begin{equation}
    \varphi\left(\mathbb{E}[x]\right) \leq \mathbb{E}[\varphi(x)] 
\end{equation}
for a continuous random variable $x$ \citep{Jensen1906}. In our case, we estimate expected values via averaging over $N_\text{e}$ samples of the random variable $p-y$, using a convex function $\varphi(p)=p^2$ for BS. This immediately yields $\Delta \text{BS} < 0$. We conclude that a skill increase through ensemble averaging is guaranteed for all convex skill scores. This has been noted  before \citep{Rougier2016}.

\subsection{ML skill decay with lead time}\label{sec:NWP_skill}

By now, we have well understood the difference in skill between our two ML models. However, we have not commented yet on their general decrease in skill as a function of lead time (e.g., \cref{fig:BSS_leadtime}). We expect skill to decrease with lead time since the forecast uncertainty of the underlying NWP model increases with lead time, as well. In this section, we investigate to what extent the decrease in ML model skill as a function of lead time can be attributed to the increasing forecast uncertainty of the underlying NWP model.

To this end, we define a surrogate variable for thunderstorm occurrence in the raw NWP output (without any ML-based corrections) and compare it to our ML predictions. As thunderstorms tend to be accompanied by heavy precipitation, radar reflectivity has been a natural proxy for thunderstorm occurrence, especially in the nowcasting community \citep{Dixon1993, Wilson1998, Turner2004}. ICON-D2-EPS generates a column-maximal radar reflectivity product, in which (synthetic) radar reflectivity is computed from simulated liquid and solid water content using a one-moment parametrization scheme \citep{Zaengl2015}.
To obtain probabilistic reflectivity forecasts which can be directly compared to SALAMA 1D output, we consider exceedance probabilities of reflectivity \citep[e.g.,][]{Theis2005, Roberts2008a}: For a given grid point of an NWP forecast, we first define its neighborhood as the set of grid points within a great-circle distance $\Delta r = \SI{15}{\kilo\meter}$, which yields $N_n=166$ neighbors. Our surrogate variable for thunderstorm occurrence in raw (deterministic) NWP data at a given grid point is then defined as the fraction of neighbors exceeding a reflectivity threshold of $\num{37}\,\text{dBZ}$. We chose this threshold after consulting previous studies in the literature which identified thunderstorms via thresholds between $\num{30}\,\text{dBZ}$ and $\num{40}\,\text{dBZ}$ \citep{Muller2003, Leinonen2022, Ortland2023}, and verifying that the following results do not change qualitatively if a different threshold within this range is chosen. The spatial threshold $\Delta r$ has been chosen to match the threshold used for the spatial aggregation of the lightning observations (\cref{sec:data}\ref{sec:lightning}), which serve as the ground truth for evaluating raw NWP skill, as well. The thusly constructed surrogate variable, henceforth called ``raw NWP'', produces probability-like output between $0$ and $1$, making a comparison to SALAMA 1D output straightforward. For ensemble forecasts, we compute exceedance probabilities for each member, and then evaluate the ensemble mean, just like for SALAMA 1D-EPS. 

In order to compare the skill of the ML-based models with those based on raw NWP output, it is crucial to take the following aspect into account: while we expect higher model reflectivities to be more frequently associated with observed thunderstorm occurrence, the exceedance probabilities are generally not well-calibrated. In turn, a calibration-sensitive skill score (e.g., BSS) displays low skill even if our surrogate variable is perfectly capable of discriminating between the two classes. Therefore, we measure skill using the resolution term \eqref{eq:res} of BSS, $\text{RES}=\sum_{i=1}^{N_\text{b}} \text{RES}_i \Delta p$,  effectively removing calibration sensitivity from BSS. Any other calibration-blind score, such as the area under the receiver operating characteristic curve, or the area under the precision--recall curve \citep{Wilks2011}, would be equally valid for the following analysis.

\begin{figure*}[htbp]
\centering
\includegraphics[width=\textwidth]{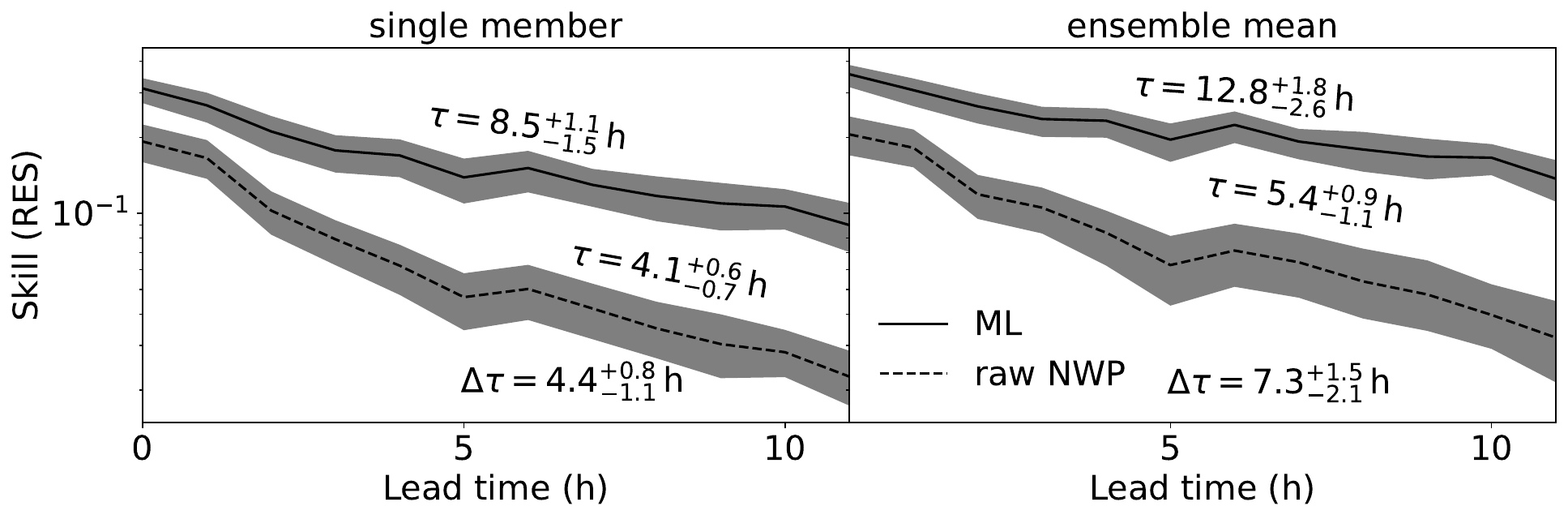}
\caption{Lead-time dependence of skill, quantified by the calibration-blind skill score RES (\cref{eq:res}) for deterministic forecasts (left panel) and ensemble-averaged forecasts (right panel). Each panel displays the results for SALAMA 1D and a simple model based on raw NWP output without any ML corrections. For each line, we fit an exponential function  $\propto\exp{(-t_\text{lead}/\tau)}$ to introduce a characteristic time scale $\tau$ of skill decay. Across all lines, the skill of ML-based forecasts decays more slowly than raw NWP forecasts, as $\Delta \tau \equiv \tau (\text{ML})-\tau (\text{raw NWP}) > 0$. Shaded bands correspond to sampling uncertainty for a symmetric \SI{90}{\percent} confidence interval. Uncertainties are obtained from \num{e4} block bootstrap resamples with day-wise block-resampling.}
\label{fig:NWP_skill}
\end{figure*}

\Cref{fig:NWP_skill} shows the skill of the SALAMA 1D models and raw NWP as a function of lead time. Raw NWP initial skill is lower than ML initial skill, which likely originates from the ML model having access to more atmospheric variables, resulting in more precise patterns of thunderstorm occurrence. Skill decreases with lead time in the case of raw NWP output, as expected from increasing NWP forecast uncertainty.  In order to compare the drop in skill quantitatively with the SALAMA 1D models, we fit exponential functions $\propto\exp{(-t_\text{lead}/\tau)}$ to each curve. The fit parameter $\tau$ then provides a characteristic time scale of skill decay---and hence, predictability---that can be compared between the individual models. It is worth noting that taking into account the entire ensemble results in longer skill decay times, no matter whether one considers the ML-based models or raw NWP output. On the other hand, the SALAMA 1D models display significantly higher skill decay times than the corresponding models based on raw NWP output. This suggests that the SALAMA 1D models' skill decay with lead time is not simply the result of increasing NWP forecast uncertainty. Instead, SALAMA 1D has learned from observations to advantageously combine multi-variable input, resulting in more persistent patterns of thunderstorm occurrence---specifically, longer-term predictable patterns---than if no observation-based postprocessing had occurred. This finding is consistent with the established understanding that the postprocessing of NWP variables with observational data leads to improved forecasts \citep{Vannitsem2021}. 

\section{Discussion and conclusion}\label{sec:conclusion}
This work aimed to contribute to the question of how ensemble NWP models can help improving thunderstorm forecasting. Specifically, we quantitatively investigated the added benefits of
ensemble-averaging, and of using an ML model instead of a traditional surrogate.

We exemplified the potential of ensemble-averaging using our recently introduced neural network model SALAMA 1D, which infers the probability of thunderstorm occurrence from vertical profiles of atmospheric variables and has been trained using forecasts of the convection-permitting NWP model ICON-D2-EPS.  We found that applying SALAMA 1D to each NWP forecast member individually and then evaluating the ensemble mean increases skill across lead times up to (at least) \SI{11}{\hour}, with an \SI{11}{\hour} ensemble forecast displaying the same skill as a \SI{5}{\hour} forecast of a single member (effectively a deterministic forecast).  We derived an analytic formula for the difference in skill (quantified by BSS) and found it to be consistent with measured difference in skill. 

A comparison with a simple model based on raw NWP output without any ML-based corrections revealed that the ML model skill decreases less quickly with lead time than the model based on raw NWP. This suggested that the decrease in ML skill with lead time is not simply a result of increasing NWP forecast uncertainty. Instead, the ML approach allows to favorably combine input from multiple atmospheric variables by systematically taking observational data into account, which is consistent with understandings in the postprocessing community.

In closing, we stress that our findings justify applying ensemble-averaging to any binary classification model of NWP ensemble forecasts that processes each member separately. As long as the correlation between the members of the underlying NWP ensemble model is sufficiently small, we expect classification skill to improve. This particularly applies to ML-based classification models whose growing role in severe weather forecasting is strengthened by our findings. Future work to improve NWP-based thunderstorm forecasting should focus on ensemble harvesting. While it is often unfeasible to increase the number of NWP model runs that make up the ensemble system, one can generate new members by including NWP variables at grid points in proximity to the location of a forecast. This ``poor man's ensemble'' in combination with the presented framework offers straightforward avenues for improved skill and decision making in the critical field of severe weather forecasting.

\clearpage
\acknowledgments
We gratefully acknowledge the computational and data resources provided through the joint high-performance data analytics (HPDA) project "terrabyte" of the DLR and the Leibniz Supercomputing Center (LRZ). C.M. carried out his contributions within the Italia--Deutschland Science--4--Services Network in Weather and Climate (IDEA-S4S; INVACODA, 4823IDEAP6). This Italian-German research network of universities, research institutes and DWD is funded by the Federal Ministry of Digital and Transport (BMDV). The authors declare that there are no conflicts of interest to disclose.

%
%
\datastatement

We aim to make the testing datasets generated in this study publicly available on Zenodo. The code for the SALAMA 1D model is released under a GPL license on GitHub (https://github.com/kvahidyou/SALAMA). The original simulation data from the ICON-D2-EPS model is accessible through the corresponding DWD database (https://www.dwd.de/EN/ourservices/pamore/pamore). Please note that the original data from the LINET lightning detection network is proprietary and cannot be shared in its raw form.

%






%



\bibliographystyle{ametsocV6}
\bibliography{main.bbl}

\end{document}